\documentclass[]{cupconf}
\usepackage{graphicx}

\newcommand{\hbeta}{\ensuremath{\mathrm{H}\beta}}
\newcommand{\mgb}{\ensuremath{\mathrm{Mg}\,b}}
\newcommand{\fe}{\ensuremath{\langle\mathrm{Fe}\rangle}}
\newcommand{\enh}{\ensuremath{\mathrm{[E/Fe]}}}
\newcommand{\afe}{\ensuremath{\mathrm{[\alpha/Fe]}}}
\newcommand{\feh}{\ensuremath{\mathrm{[Fe/H]}}}

\newcommand{\z}{\ensuremath{\mathrm{[Z/H]}}}

\title[Early-type galaxy abundances]{Stellar abundances of early-type galaxies}
      
\author[S.C.~Trager]{S. C. Trager} 
\affiliation{Kapteyn Astronomical Institute, University of Groningen\\
  Postbus 800, NL-9700 AV Groningen, Netherlands}

\begin{document}
\maketitle

\begin{abstract}
It is currently impossible to determine the abundances of the stellar
populations star-by-star in dense stellar systems more distant than a
few megaparsecs.  Therefore, methods to analyse the composite light of
stellar systems are required.  I review recent progress in determining
the abundances and abundance ratios of early-type galaxies.  I begin
with `direct' abundance measurements: colour--magnitude diagrams of
and planetary nebula in nearby early-type galaxies.  I then give an
overview of `indirect' abundance measurements: inferences from stellar
population models, with an emphasis on cross-checks with `direct'
methods.  I explore the variations of early-type galaxy abundances as
a function of mass, age, and environment in the local Universe.  I
conclude with a list of continuing difficulties in the modelling that
complicate the interpretation of integrated spectra and I look ahead
to new methods and new observations.
\end{abstract}

\firstsection
\section{Introduction}

The spheroidal components of galaxies -- elliptical and lenticular
(S0) galaxies and the bulges of spirals -- contain at least half of
the present-day stellar mass in the local Universe (Schechter \&
Dressler 1987; Fukugita, Hogan \& Peebles 1998).  An understanding the
formation and evolution of these objects is therefore necessary in
order to understand the dominant galaxy types in the local Universe.

The abundances of stars in these objects give us direct handles on
their formation and evolution.  The gross metallicity of a galaxy
tells us about its overall chemical evolution.  Abundances of specific
elements are much more useful, as these give direct evidence to the
nucleosynthetic processes that occurred during the formation of that
star (and by extension, of the stellar population and even galaxy).
As a example, the ratio \afe\ is crudely an indicator of the ratio of
SNe II, which result from the explosion of massive stars and produce
the $\alpha$-elements, to SNe Ia, which result from the explosion of
low- to intermediate-mass stars and produce most of the Fe-peak
elements.  This leads to \afe\ being commonly used as a tracer of the
timescale of star formation, as it will decrease with increasing
length of star formation as SNe Ia become more important in the
nucleosynthesis in a stellar population (e.g., Worthey, Faber \&
Gonz\'alez 1992; Greggio 1997; Trager et al.\ 2000b; Thomas et
al.\ 2005; below).  This is an oversimplification, of course, as can
be seen in papers by Matteucci and Pipino \& Matteucci (both in this
volume).

Unfortunately, the Universe has made measuring stellar abundances in
these objects difficult for us.  Spheroids of galaxies are very dense
and the crowding makes it impossible to make accurate photometric
measurements of individual stars closer than one effective radius away
from their centres in all spheroidal objects but our own Galactic
Bulge (and in that case, dust presents a formidable barrier), even
with the \textsl{Hubble Space Telescope} (\textsl{HST}).  Attempting
to determine abundances of individual stars from high-resolution
spectra, as we do in the Milky Way, is accordingly impossible beyond
the Local Group with the current generation of 8--10m ground-based
telescopes and the 2.5m \textsl{HST}.  The best we can hope for is to
measure the colour--magnitude diagrams of the outer regions of these
galaxies and infer their abundances from the distribution of stars on
the giant branches, attempt to determine abundances from planetary
nebulae -- which stand out from the crowd as emission-line objects --
and finally to infer abundances from the integrated light of the
galaxies.  Each of these methods will be discussed below.

In the following, I refer to as `direct' methods those techniques that
determine abundances from resolved stellar populations.
Colour--magnitude diagrams fall into this category, even though
abundances of \emph{individual} stars may be poorly determined, as do
abundances from analyses of planetary nebula spectra.  I refer to as
`indirect' methods those techniques that determine abundances from
unresolved stellar populations, in particular analysis of
absorption-line strengths.  I discuss direct abundance measurements in
\S\ref{sec:direct} and indirect abundance measurements in
\S\ref{sec:indirect}.  I summarise the results in \S\ref{sec:summary}.

To conclude this section, I mention that there are other methods for
determining the abundances of early-type galaxies, including the
analysis of their globular cluster systems and determinations of the
abundances of X-ray emitting gas.  However, for reasons of space and
of complications in relating these determinations to the stellar
population of the host galaxy, I will not discuss them here.  For the
former, the interested reader can refer to Trager (2004) and Brodie \&
Strader (2006); for the later, see, e.g., Humphrey \& Buote (2006).
Finally, Jablonka has covered the subject of bulges of spiral galaxies
elsewhere in this volume.

\section{Direct abundance measurements}
\label{sec:direct}

I begin with an overview of determinations of stellar population
abundances of nearby galaxies using direct abundance measurements:
colour--magnitude diagrams and planetary nebulae.

\subsection{Abundances from colour--magnitude diagrams}

\begin{figure}
  \includegraphics[width=\textwidth]{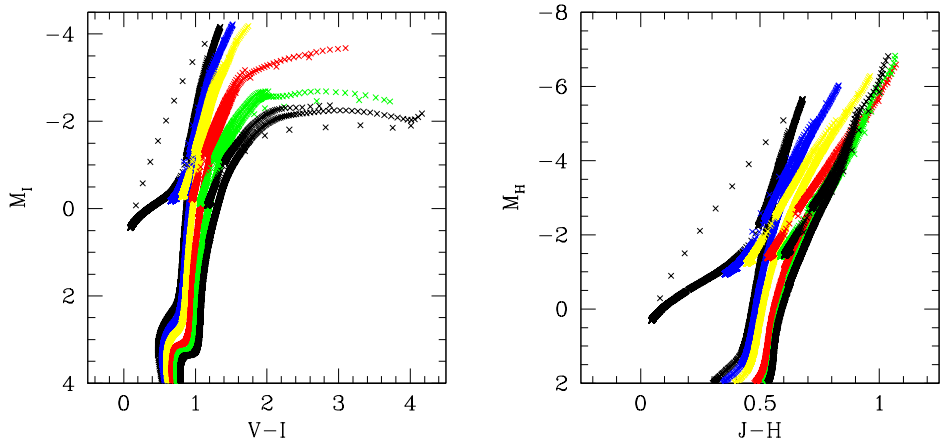}
  \caption{The giant branches of old (12 Gyr) stellar populations from
    the BaSTI (Pietrinferni et al.\ 2004, 2006) isochrone collection,
    in the optical (left) and near-infrared (right).  These isochrones
    include the horizontal branch and asymptotic giant branch phases.
    Metallicities, from left to right in both panels, are
    $\feh=-2.27$, $-1.49$, $-0.96$, $-0.35$, $0.06$, and $0.40$.}
  \label{fig:basti}
\end{figure}

Currently, at least seven (non-dwarf) elliptical and S0 galaxies have
observed colour--magnitude diagrams (CMDs): M32
(Grillmair et al.\ 1996), NGC 5128 (Rejkuba et al.\ 2005 and
references therein), NGC 3379 (Gregg et al.\ 2004), NGC 3115, NGC
5102, NGC 404 (Schulte-Ladbeck et al.\ 2003), and Maffei 1 (Davidge \&
van den Bergh 2001).  For all of these galaxies, CMDs have required
\textsl{HST} imaging in the optical or near-infrared (NIR), except for Maffei
1, which was observed in the NIR with the CFHT adaptive optics system.
However, in only three cases -- M32, NGC 3379, NGC 5128 -- have the
CMDs reached \emph{significantly below} the tip of the red giant
branch (TRGB), required for reasonable abundance measurements
(Fig.~\ref{fig:basti}, left).  In the other cases, only the asymptotic
giant branch was detected (Maffei 1) or only stars less than one
magnitude below the tip of the red giant branch were detected.  In NGC
3779, although the NICMOS F110W ($J$) and F160W ($H$) measurements
extend to three magnitudes below the tip of the asymptotic giant
branch (AGB), the tight spacing of isochrones of different
metallicities and the co-mingling of the AGB and RGB over this
magnitude range makes precision estimates of stellar abundances very
difficult, and only ranges of acceptable metallicities can be inferred
rather than accurate metallicity \emph{distributions}
(Fig.~\ref{fig:basti}, right; Gregg et al.\ 2003).

\begin{figure}
  \includegraphics[width=0.45\textwidth]{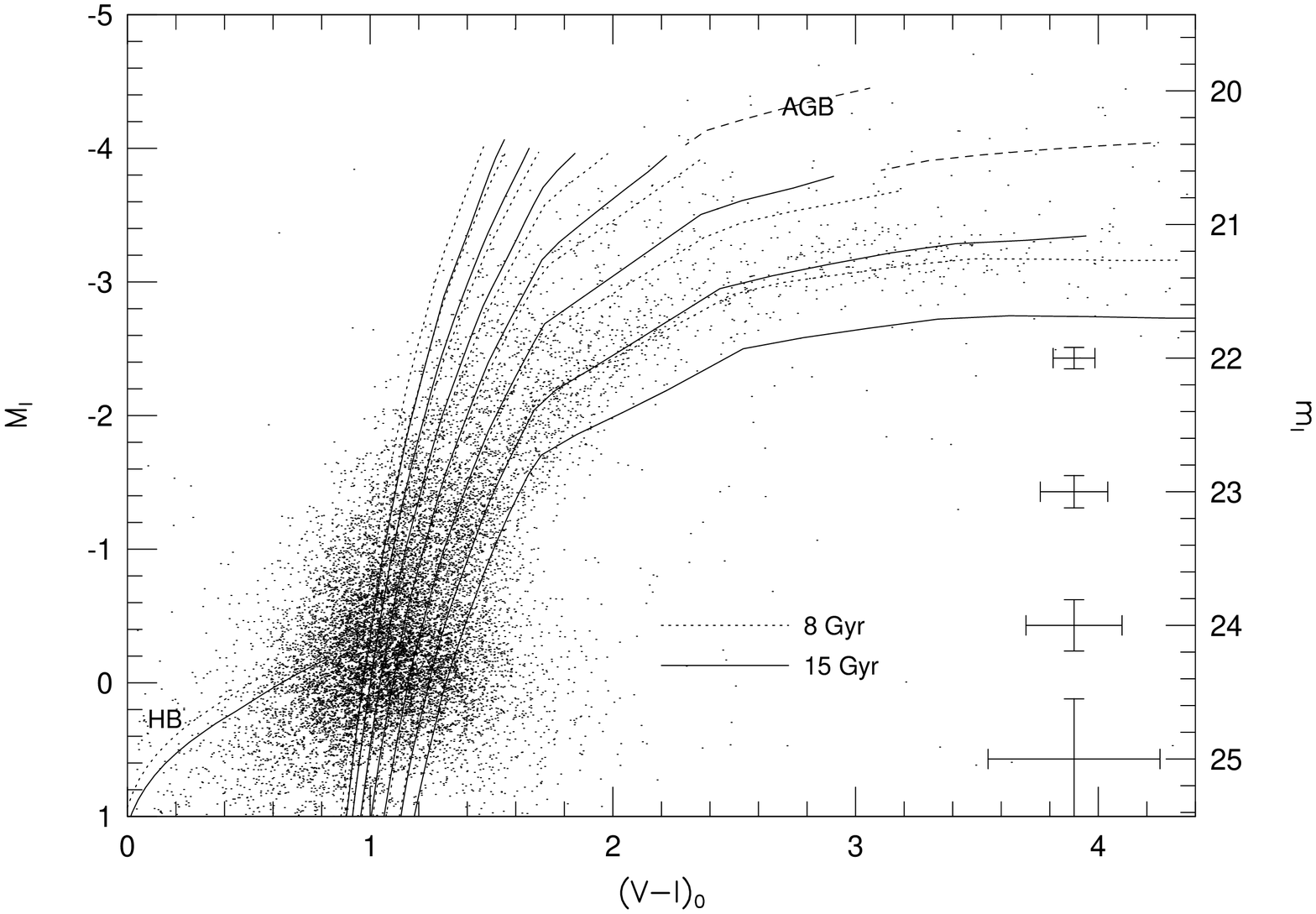}
  \hfill
  \includegraphics[width=0.45\textwidth]{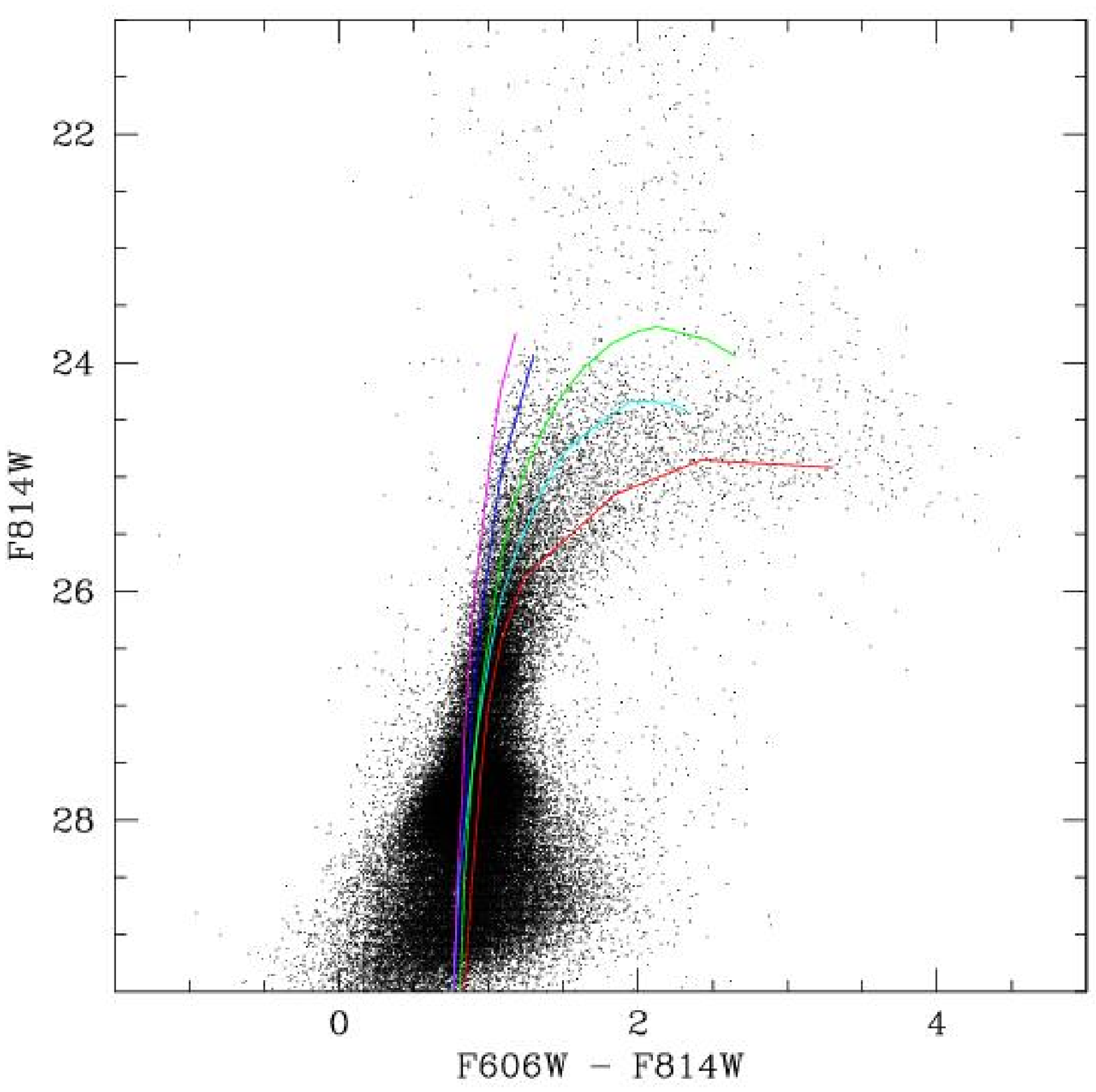}
  \caption{Colour--magnitude diagrams of M32 (left; Grillmair et
    al.\ 1996) and NGC 5128 (right; Rejkuba et al.\ 2005).  In the
    left panel, isochrones from Worthey (1994) have been over-plotted
    with metallicities between $\feh=-1.2$ to 0 dex in 0.2-steps for
    ages of 10 Gyr (dotted lines) and 15 Gyr (solid lines).  Fiducial
    globular cluster colour--magnitude sequences have been over-plotted
    in the right panel: NGC 6341, NGC 6752, 47 Tuc, NGC 5927, and NGC
    6528 (from left to right).}
  \label{fig:cmds}
\end{figure}

\begin{figure}
  \includegraphics[width=0.45\textwidth]{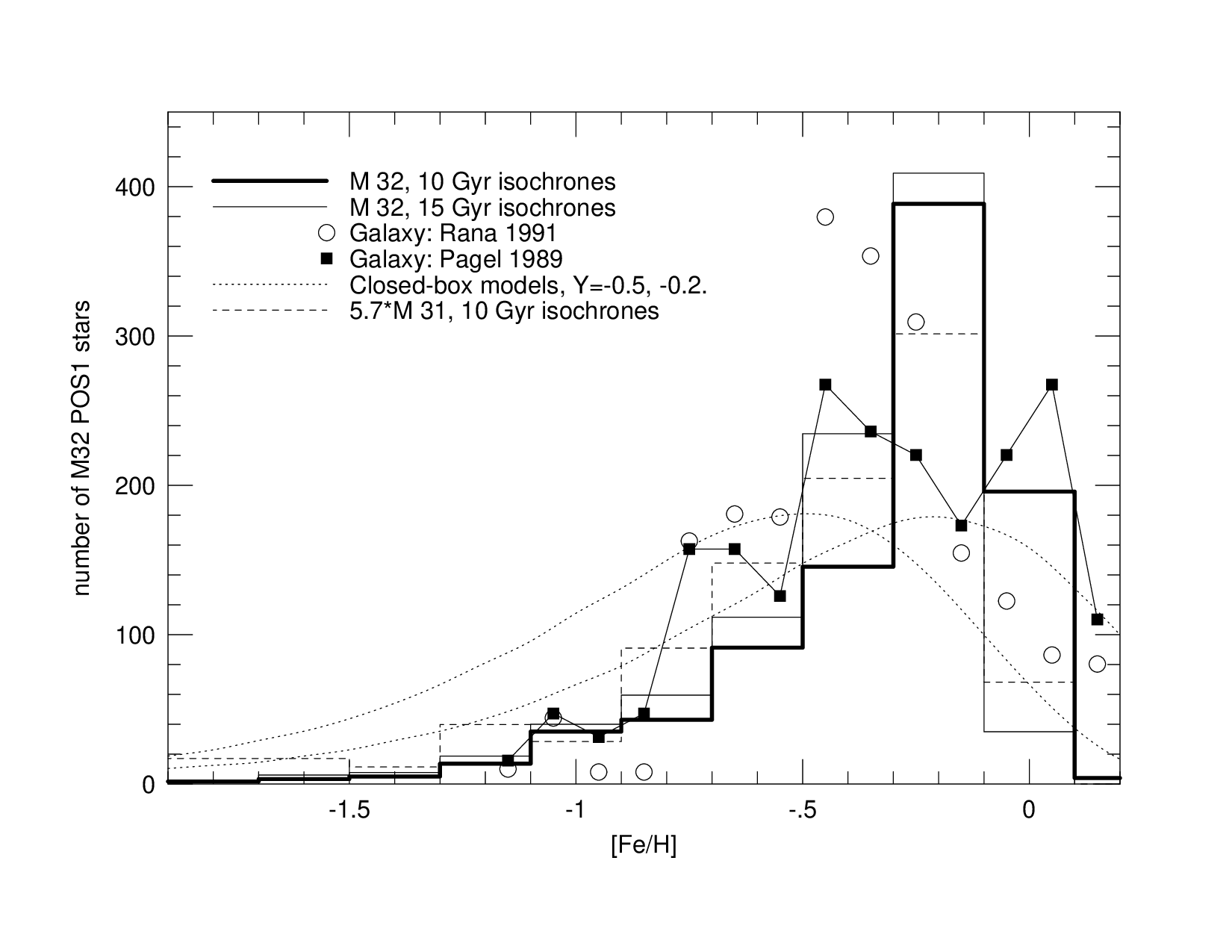}
  \hfill
  \includegraphics[width=0.45\textwidth]{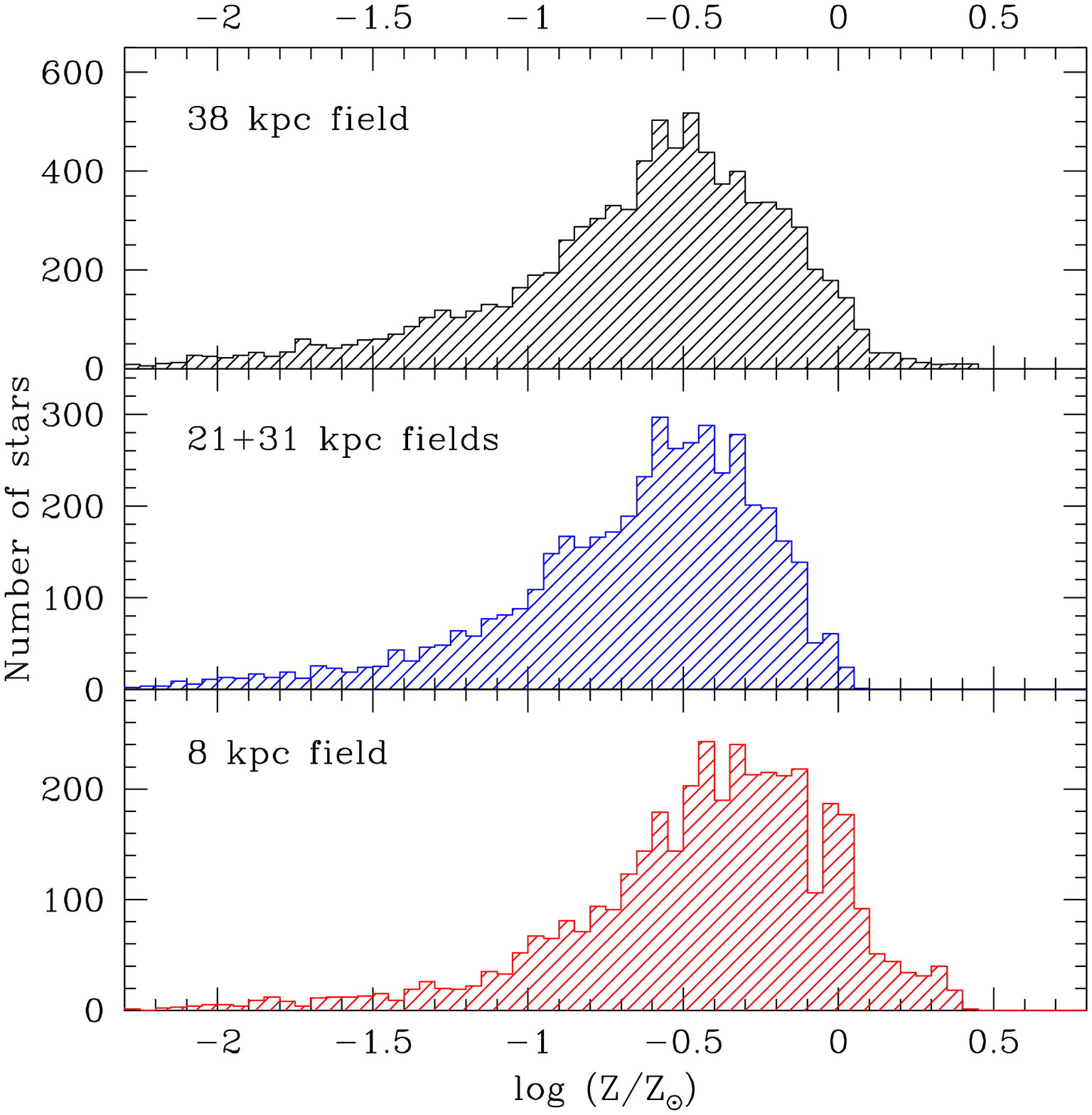}
  \caption{Metallicity distribution functions for M32 (left; Grillmair
    et al.\ 1996) and NGC 5128 (right; Rejkuba et al.\ 2005) inferred
    from comparing the distribution of RGB stars to isochrones.  For
    M32, the isochrones used were 10 or 15 Gyr old (see legend); for
    NGC 5128, the isochrones used were 12 Gyr old.  For M32,
    closed-box chemical evolution (`Simple') models have been
    over-plotted with two different choices of the yield.}
  \label{fig:mdfs}
\end{figure}

In M32, Grillmair et al.\ (1996) obtained a CMD at a distance of
$r\sim3r_e$ with WFPC2 on \textsl{HST}.  Because M31 is superimposed on M32,
the resulting CMD was statistically cleaned of M31 stars by using a
nearby M31 disk field.  The resulting CMD reaches a depth just below
the `red clump', the pile-up of red horizontal branch (helium burning)
stars about four magnitudes below the TRGB (Fig.~\ref{fig:cmds},
left).  There was no evidence of a blue horizontal branch (BHB) in
this CMD, although the limiting magnitude in $V$ prevented ruling out
its presence completely.  Their absence has been confirmed in deeper
WFPC2 (Worthey et al., in prep.) and ACS/HRC (Lauer et al., in prep.)
observations.  This is a crucial result, as BHB stars trace the most
metal-poor stars in our Galaxy: at metallicities $\feh<1.5$,
helium-burning stars are dominantly BHB stars in Galactic globular
clusters.  Their lack in M32 suggests that this galaxy is deficient in
metal-poor stars, which is confirmed by its metallicity distribution
inferred from the distribution of RGB stars
(Fig.~\ref{fig:mdfs}, left).  M32 therefore has a `G-dwarf
problem' (Worthey, Dorman \& Jones 1996), for which the usual
suggested solution is pre-enrichment of the galaxy (Tinsley 1980).

In NGC 5128, Rejkuba et al.\ (2005, following earlier studies of inner
fields by Harris \& Harris 2000, 2002) obtained a CMD at a distance of
$r\sim7r_e$ with ACS/WFC on \textsl{HST}.  As in M32, Rejkuba et al.\ obtained
a CMD down to the red clump (Fig.~\ref{fig:cmds}, right) and found no
evidence of a BHB.  Again, it is difficult to rule out their presence
completely, but the metallicity distribution inferred from the
distribution of RGB stars (Fig.~\ref{fig:mdfs}, right) suggests that
there are few (but some) very-metal-poor stars in NGC 5128.

By counting stars on the RGB between isochrones of different
metallicities (but typically constant age), a metallicity distribution
function (MDF) can be determined.  MDFs from M32 and NGC 5128 are
shown in Figure~\ref{fig:mdfs}.  Both galaxies have a deficiency of
metal-poor stars compared to a closed-box (`Simple') chemical
evolution model, as discussed above.  M32 has a sharp peak in its MDF
at $\feh=-0.25$ (using isochrones from Worthey 1994).  In
\S\ref{sec:spresults} below I will compare this inferred MDF with
results from stellar population analyses.  The MDF of NGC 5128 has a
significant radial gradient and peaks at $\feh\approx-0.6$ to
$\approx-0.3$, from the outermost to innermost fields (using
isochrones from VandenBerg et al.\ 2000, re-calibrated to match
Galactic globular clusters; see Rejkuba et al.\ 2005 for more
details).  Note that the lack of high-metallicity stars in the 21--31
kpc field of NGC 5128 is a selection effect due to insufficiently deep
$V$-band observations, so the shape of this MDF cannot be fairly
compared to the inner and outer fields.  Even so, in both galaxies,
there are stars of \emph{at least} solar metallicity, and there is
evidence for super-metal-rich stars, particularly in the innermost
field.

There are certain cautions to be kept in mind when considering
CMD-inferred MDFs.  First, old, single-age stellar populations are
\emph{assumed} in order to reliably compute these distributions.  This
is due to the lack of age information in insufficiently-deep CMDs:
without CMDs that reach down below the main-sequence turn-off(s), the
age is unknown (although the presence of bright AGB stars can indicate
intermediate-aged populations).  A shift, or worse, a distribution in
stellar population age can change both the \emph{mean and shape} of
the inferred MDF.  This can be seen by comparing the MDFs of M32
inferred from the 10 and 15 Gyr old isochrones in
Figure~\ref{fig:mdfs} and imaging a mixture of these populations.
Second, the assumed abundance \emph{pattern} (as parametrised by, say,
\afe) can also alter both the mean and shape of the inferred MDF.
This is because changes in the abundance pattern can alter the
temperature of the stars on the giant branch (e.g., Salaris, Chieffi
\& Straniero 1993).  On the other hand, this may not be a significant
problem for these two galaxies: for M32, because its abundance pattern
is very close to that of the Sun (Worthey 2004), and the isochrones
used to infer its MDF have scaled-solar composition (Worthey 1994);
and for NGC 5128, because while the galaxy likely has $\afe>0$ because
it is a giant elliptical (see below), the isochrones used to infer its
MDFs were $\alpha$-enriched (Rejkuba et al.\ 2005).

\subsection{Abundances from planetary nebulae}

Planetary nebulae (PNe) are the second-to-last phase of the life of
all low- and intermediate-mass stars.  Although this phase is quite
short, the large flux of stars through it means that many PNe are
visible at any given time.  Because this phase is so short, their
density is much lower than the field stars from which they evolved,
and therefore they are much less crowded on the sky.  PNe are
emission-line objects and are fairly easy to distinguish from the
absorption-line background of their hosts.  Using standard
emission-line techniques it is possible to determine elemental
abundances directly for each PNe.

However, PNe are quite faint.  Special wide-field spectroscopy
techniques have been developed to determine the kinematics of PNe
systems around galaxies (e.g., Douglas et al.\ 2002), which utilise a
narrow-band region around the [O\textsc{iii}]$\lambda5007$ line.  But
these techniques do not (yet) allow for abundance determinations, as
these require a larger wavelength region: oxygen abundances require
measurements of the [O\textsc{iii}]$\lambda4363,5007$ and
[S\textsc{ii}]$\lambda6716,6731$ lines.  Moreover, the
[O\textsc{iii}]$\lambda4363$ line is very weak.  Unfortunately,
without this line, the electron temperature $T_e$ cannot be measured
directly, and strong-line techniques (such as those used for
H\textsc{ii} regions) cannot be used for PNe because the temperature
of the central stars vary significantly from object to object, and
photoionization models do not give unique determinations
(Stasi{\'n}ska et al.\ 2006).  On the other hand, unlike in
H\textsc{ii} regions, the [O\textsc{iii}]$\lambda4363$ line is still
visible in metal-rich PNe because of the high density of the nebulae
and the high temperatures of the central stars (Stasi{\'n}ska et
al.\ 2005).  With large telescopes ($\ge4$m) and sensitive
spectrographs, bright PNe in early-type galaxies become reasonable
tracer objects for the abundance patterns of their hosts.

Many early-type galaxies host well-known PNe systems, although for
only a handful have abundance measurements been attempted.  Here I
will concentrate on the three cases that I know of that have
independent abundance determinations from other methods: M32
(Stasi{\'n}ska, Richer \& McCall 1998), NGC 5128 (Walsh et al.\ 1999,
2006), and NGC 4697 (M\'endez et al.\ 2005).  Of these, only M32 and
NGC 5128 have PNe with detected [O\textsc{iii}]$\lambda4363$ lines: 9
PNe in M32 (Stasi{\'n}ska et al.\ 1998) and 2 (so far) in NGC 5128
(Walsh et al.\ 2006).  None of the PNe in NGC 4697 show
[O\textsc{iii}]$\lambda4363$ individually, although it is detected in
an average spectra (M\'endez et al.\ 2005).  I will therefore
(reluctantly) no longer consider it here, as the $T_e$ of the
individual PNe cannot be determined and therefore their abundances are
likely to be very uncertain.  In M32, Stasi{\'n}ska et al.\ (1998)
found $\langle[\mathrm{O/H}]\rangle=-0.5$, in reasonable agreement
with and at projected galactocentric distances similar to those of the
Grillmair et al.\ (1996) CMD study.  Walsh et al.\ (1999) find a
similar average [O/H] abundance ratio for NGC 5128 in a field close to
the centre, although none of these PNe had detected
[O\textsc{iii}]$\lambda4363$.  Walsh et al.\ (2006) appear to confirm
this result using two PNe with $T_e$ estimates and also find evidence
for two distinct populations of PNe in NGC 5128, a `bulge' and a
`halo' population.

Abundances from PNe appear to be a rich area for further study, but
the results of M\'endez et al.\ (2005) and Walsh et al.\ (2006)
suggest that perhaps the current generation of telescopes are at their
limits even for PNe in galaxies as close as NGC 5128 (3.8 Mpc; Rejkuba
et al.\ 2005).

\section{Indirect abundance measurements}
\label{sec:indirect}

Given the difficulties of determining the abundances and abundance
distributions star-by-star in even nearby early-type galaxies, another
method is required.  Analysis of the integrated light of these
galaxies allows us to determine abundances and abundance patterns in
entire stellar populations at over a significant fraction of the
history of the Universe (currently out to $z\sim0.8$: J{\o}rgensen et
al.\ 2005).

\subsection{Stellar population models}

Here I give a short overview of the method, which involves applying
stellar populations to strengths of absorption lines tuned to decouple
age and metallicity (and abundance ratios) in the spectra.  A more
complete description can be found in Trager (2004); for details the
reader is referred to Trager et al.\ (2000a), with updates in Trager,
Faber \& Dressler (2006).  An unfortunate drawback of this method at
present is its inability to determine MDFs; only mean quantities
(weighted in a peculiar way described below) can be determined.


At first glance, it might seem natural to use optical broad-band
colours to determine rough abundances of galaxies: metal-poor stellar
systems are blue and metal-rich ones are red.  This is of course
because the increasing opacity from metals removes light from the blue
and moves it into the red, particularly on the RGB, which provides at
least half of the optical light of galaxies.  Unfortunately, age has
the same effect: older populations are cooler because they have lower
mass on the RGB and are therefore redder.  One might expect to improve
the situation by using metallic absorption lines like the
\mgb\ feature, the MgH triplet (Mg$_2$), or the Fe lines at
5270\AA\ and 5335\AA, which would allow a direct measurement of the
metallicity.  Unfortunately, these lines have the same problem as
broad-band colours: they are formed in the atmospheres of the cool RGB
stars and are sensitive to their temperatures -- and are therefore
also subject to variations in stellar population age.  This
age--metallicity degeneracy (e.g., O'Connell 1980) was finally broken
by Worthey (1994) and Buzzoni, Mantegazzi \& Gariboldi (1994), who
independently showed that a plot of Balmer-line strength (such as
\hbeta) as a function of metal-line strength could allow an
independent measurement of stellar population age and metallicity (a
result first demonstrated by Rabin 1982).  This is because the
temperature of the main-sequence turn-off (MSTO) of a stellar
population is more sensitive to age than metallicity, and the Balmer
lines of hydrogen are non-linearly sensitive to the temperature of the
MSTO.

Stellar population models can then be built that predict the metal-
and Balmer-line strengths of stellar populations (e.g., Worthey 1994).
Stellar interior calculations, in the form of isochrones, are combined
with stellar fluxes, either empirically or theoretically determined,
and stellar absorption-line strengths, which are almost always
empirically determined (but interpolation methods vary) to produce
predicted line strengths as a function of stellar population age and
composition.  Modern models (e.g., Trager et al. 2000a; Thomas,
Maraston \& Bender 2003) now allow for variations in abundance ratios
like \afe, as models based on line strengths of solar-neighbourhood
stars produce (for example) \mgb\ line strengths too weak for a given
\fe\ line strength compared to giant elliptical galaxies (see, e.g.,
Peterson 1976; O'Connell 1980; Peletier 1989; Worthey et al.\ 1992;
Fig.~\ref{fig:hbmgfe}).

\subsection{Stellar population analyses}
\label{sec:spresults}

\begin{figure}
  \includegraphics[width=\textwidth]{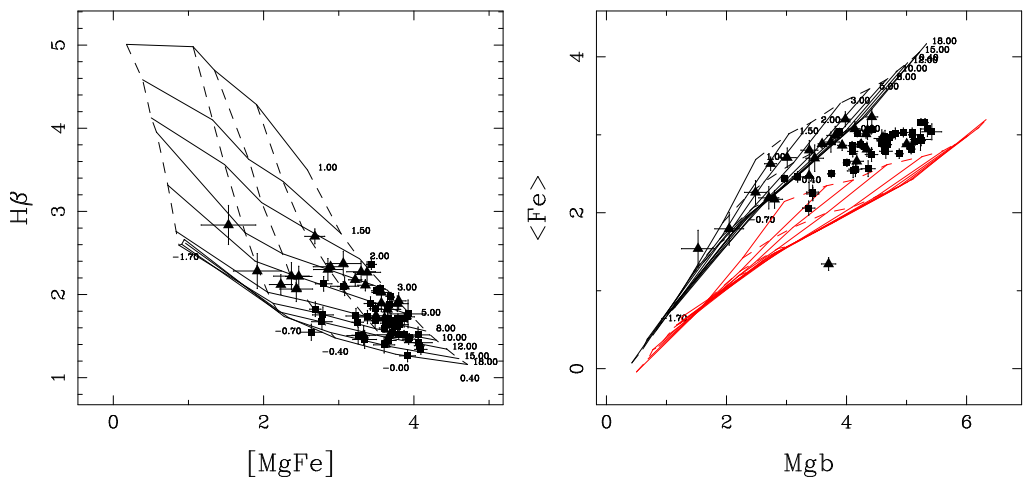}
  \caption{Stellar populations of nearby early-type (elliptical and
    S0) galaxies.  Elliptical galaxies in the field are taken from
    Gonz\'alez (1993), S0s in the field from Fisher, Franx \&
    Illingworth (1996), and elliptical and S0s in the Fornax Cluster
    from Kuntschner (2000).  Ellipticals are squares and S0s are
    triangles.  Model grids from Bruzual \& Charlot (2003) as modified
    by Trager et al.\ (2006; cf.\ Serra \& Trager 2006) for
    $\afe\neq0$ are over-plotted.  Solid lines have constant age;
    dashed lines have constant metallicity.  In the left panel,
    $[\mathrm{MgFe}]=\sqrt{\mgb\times\fe}$ is a metallicity indicator
    reasonably free of non-solar-abundance-ratio effects (Thomas et
    al.\ 2003).  This means that SSP-equivalent age and metallicity
    can be read from this panel.  In the right panel, model grids have
    $\afe=0$, $+0.3$ from left to right.  This panel is used to
    estimate $\afe$.}
  \label{fig:hbmgfe}
\end{figure}

Once stellar population models are available, stellar population ages,
metallicities, and abundance ratios can be read off of diagrams like
Figure~\ref{fig:hbmgfe}.  (Note however that stellar population
parameters are actually inferred from $\chi^2$ minimisation of the
model line strengths against the observed line strengths.)  All
early-type galaxies span a range in metallicity and age (Trager et
al.\ 2000b), with the mean age being older in clusters than in the
field (Thomas et al.\ 2005).  Early-type galaxies in all environments
typically have $-0.5<\z_{\mathrm{SSP}}<+0.5$ and
$0<\afe_{\mathrm{SSP}}<+0.4$ (depending exactly on the stellar
population model used).  There appears to be no significant difference
in metallicity between different environments (Thomas et al.\ 2005;
S\'anchez-Bl\'azquez et al.\ 2006), although possible small
differences in [C/Fe] or [N/Fe] are possible (S\'anchez-Bl\'azquez et
al.\ 2006).


An important caveat must be understood when reading diagrams like
Figure~\ref{fig:hbmgfe}.  The ages and compositions inferred from
these diagrams are those of the equivalent single stellar population
(we call them the `SSP-equivalent' parameters).  That is, these are
the ages and compositions that the objects would have \emph{if} they
were composed solely of a single population formed in a single burst
at the SSP-equivalent age with a chemical composition given by the
SSP-equivalent metallicity and abundance ratio(s).  This means that
the composite populations of real galaxies are treated in terms of
their \emph{line-strength-weighted mean} population parameters.  In
the case of M32, we can study what effect the MDF has on the inferred
SSP-equivalent metallicity.  The light-weighted mean metallicity from
the MDF is $\feh=-0.25$, as can be inferred from Figure~\ref{fig:mdfs}
(left).  Using absorption-line measurement extrapolated (slightly) to
precisely the same position in the galaxy, $\feh_{\mathrm{SSP}}=-0.32$
(Trager et al.\ 2000b).  This excellent agreement between the
metallicities inferred from the integrated light and the CMD is only
possible in the absence of a metal-poor tail in the MDF, as hot stars
corrupt the abundances inferred from line-strengths (Trager et
al.\ 2005).  More extensive analyses using synthetic star formation
histories and chemical evolution models shows that the inferred
SSP-equivalent metallicities and \afe\ ratios are very close to their
light-weighted or even mass-weighted mean values (Serra \& Trager
2006; Trager \& Somerville, in prep.).  This is certainly \emph{not}
the case for the SSP-equivalent ages, which can be easily skewed to
young ages by the presence of very small amounts of young stars.

Using long-slit spectra of M32 and sophisticated stellar population
models that include in a limited way the variation of individual
elements as a function of Fe abundance, Worthey (2004) finds
$\feh=+0.02$, $\mathrm{[C/Fe]}=+0.08$, $\mathrm{[N/Fe]}=-0.13$, and
$\mathrm{[Mg/Fe]}=-0.18$ near the nucleus.  This confirms the
generally-held notion that M32 has an abundance pattern very much like
that of the Sun (O'Connell 1980; Grillmair et al.\ 1996; Trager et
al.\ 2000a).  The inferred nitrogen abundance appears to be too low
compared to the nitrogen abundances of the PNe (Stasi{\'n}ska et
al.\ 1998), but this could be due to self-enrichment of nitrogen in
the PNe.

\begin{figure}
  \includegraphics[width=\textwidth]{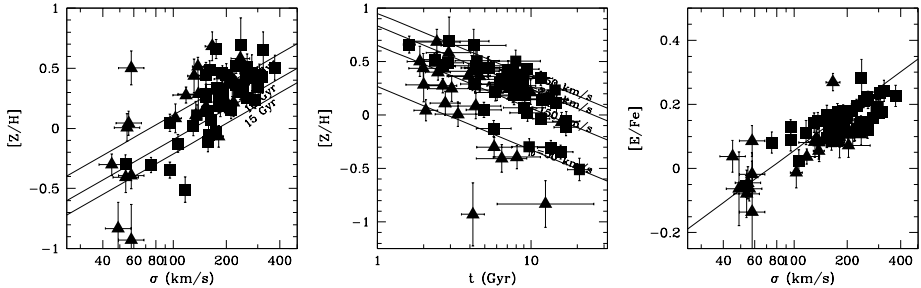}
  \caption{The variation of stellar population parameters with age and
    velocity dispersion ($\sigma$).  From left to right: the velocity
    dispersion--metallicity plane; the age--metallicity plane; and the
    velocity dispersion--abundance ratio plane.  Points are as in
    Fig.~\ref{fig:hbmgfe}.  Stellar population parameters inferred
    from models of Worthey (1994) altered as described in Trager et
    al.\ (2006).  Lines in the left and middle panels are projections
    of the $Z$-plane (Trager et al.\ 2000b).  In the left panel, the
    lines correspond to ages of 5, 10, and 15 Gyr from top to bottom;
    in the middle panel, the lines correspond to $\sigma=50$, 150,
    250, and $350\,\mathrm{km\,s^{-1}}$.  The line in the right panel
    is the $\enh$--$\sigma$ relation of Trager et al.\ (2000b).}
  \label{fig:sigmarels}
\end{figure}

With this caveat in mind, the variations of stellar population
parameters with velocity dispersion (as a tracer of the mass) can be
examined.  The left panel of Figure~\ref{fig:sigmarels} shows the
mass--metallicity relation for field galaxies.  That a narrower
mass--metallicity relation is not apparent in this plot is due to the
\emph{real} anti-correlation between SSP-equivalent age and
metallicity of galaxies in this sample (middle panel): older field
galaxies have lower metallicities at fixed velocity dispersion.  The
dispersion in ages causes a broadening of the mass--metallicity
relation.  A narrow mass--metallicity relation is however seen clearly
in cluster samples (e.g., Nelan et al.\ 2005; Trager et al.\ 2006),
because these samples have small age spreads.  At the same time, there
is a clear mass--\afe\ relation in all environments, which is only
weakly dependent on (Thomas et al.\ 2005; Nelan et al.\ 2005) or
independent of (Trager et al.\ 2000b) SSP-equivalent age.  The
physical mechanisms driving these relations are likely to be a
combination of rapid star formation, so that SNe II products dominate
over SNe Ia products, and metal-enriched outflows, in which high-mass
galaxies retain their SNe II products -- which dominate the overall
metallicity of the systems -- more easily than low-mass galaxies
(Trager et al.\ 2000b; and see Matteucci and Pipino \& Matteucci, this
volume).

\subsection{Continuing annoyances}

Nothing is easy, of course.  Besides the effects of composite
populations on the inferred stellar population parameters and the fact
that MDFs cannot (currently) be determined from the integrated light,
two major annoyances persist when trying to infer abundances from
absorption-line measurements.  The first is the unknown oxygen
abundances of early-type galaxies.  The oxygen abundance controls the
temperature of the MSTO (Salaris \& Weiss 1998), and so the inferred
age -- and therefore metallicity -- is dependent on the oxygen
abundance.  It is not yet clear if a good tracer of oxygen abundance
is available in the optical (but see below).  The second is the
hot-star content of these galaxies.  Trager et al.\ (2005) have shown
that these stars, such as BHB and blue straggler stars, not only
corrupt age measurements for the oldest objects (e.g., Maraston \&
Thomas 2000) but also corrupt abundance measurements.  Correcting for
this effect requires excellent models and data with sensitivity in the
blue, but this has been done for very few galaxies.

\section{Summary}
\label{sec:summary}

Early-type galaxies are dominantly metal-rich, of typically solar
metallicity, and usually enriched in SNe II products.  Their
compositions -- both bulk metallicity and abundance ratios -- depend
on their mass, and possibly on their environment.  Their MDFs appear
to violate `Simple' chemical evolution models and therefore these
galaxies suffer like the Milky Way from the G-dwarf problem.  They
appear to have complex star formation histories: galactic winds and
rapid star formation have apparently played crucial roles in their
chemical evolution; and there is mounting evidence for recent star
formation, suggesting perhaps recent accretion or mergers playing a
role.  Better understanding of their nucleosynthetic properties would
be very helpful to disentangle these histories.

The future for the understanding the nucleosyntetic histories of
early-type galaxies is bright.  Serven, Worthey \& Briley (2005) have
developed 54 new indices over a wide wavelength range in the optical
that will help to determine C, N, O, Na, Mg, Al, Si, Ca, Sc, Ti, V,
Cr, Mn, Fe, Co, Ni, Sr, and Ba abundances.  A significant
computational effort is underway by Worthey and collaborators
(including the author) to produce fully self-consistent stellar
interior and atmosphere models (and emergent spectra) to take
advantage of these indices and their predecessors, the Lick/IDS
indices.  At the same time, line-strength measurements are being
pursued both with two-dimensional spectrographs like \textsc{Sauron}
to study local galaxies (e.g., Kuntschner et al.\ 2006) and with
multislit spectrographs on large telescopes to study very distant ones
(e.g., J{\o}rgensen et al.\ 2005).  New observations of nearby
early-type with ACS on \textsl{HST} (such as the Lauer et al.\ and
Worthey et al.\ studies of M32) will greatly improve our knowledge of
their MDFs (and ages).  Hopefully the next generation of large
telescopes will enable star-by-star chemical studies of the nearest
elliptical galaxies!

\smallskip

I thank the SOC and LOC, particularly Garik Israelian, for the kind
invitation to review this subject and for their financial support that
enabled my participation.  I wish also to thank my collaborators, Alan
Dressler, Sandy Faber, Wendy Freedman, Carl Grillmair, Tod Lauer, Ken
Mighell, Paolo Serra, Rachel Somerville, and Guy Worthey for allowing
me to present results from on-going studies.  Thanks to Carl Grillmair
and Marina Rejkuba for allowing me to reprint two figures each.
Finally, I would like to thank Grazyna Stasi{\'n}ska for a very
helpful discussion about the use of planetary nebulae as abundance
indicators, Patricia S\'anchez-Bl\'azquez for several helpful
conversations, and Marina Rejkuba for reading a draft of the
manuscript.



\begin{thebibliography}{99}

\bibitem[]{Bender, Burstein, & Faber1992} Bender, R., Burstein, D., and 
Faber, S.~M.\ (1992). \textit{ApJ}, \textbf{399}, 462.  

\bibitem[]{Brodie & Strader2006} Brodie, J.~P.~and Strader,
  J.\ (2006).  \textit{ARA\&A}, \textbf{44}, 193.

\bibitem[]{Bruzual & Charlot2003} Bruzual, G.~and Charlot,
  S.\ (2003).  \textit{MNRAS}, \textbf{344}, 1000.

\bibitem[]{Buzzoni, Mantegazza, & Gariboldi1994} Buzzoni, A.,
  Mantegazza, L., and Gariboldi, G.\ (1994). \textit{AJ},
  \textbf{107}, 513.

\bibitem[]{Davidge & van den Bergh2001} Davidge, T.~J.~and van den
  Bergh, S.\ (2001). \textit{ApJ}, \textbf{553}, L133.

\bibitem[]{Douglas et al.2002} Douglas, N.~G., Arnaboldi, M., Freeman,
  K.~C., et al.\ (2002). \textit{PASP}, \textbf{114}, 1234.

\bibitem[]{Fisher, Franx, & Illingworth1996} Fisher, D., Franx, M.,
  and Illingworth, G.\ (1996). \textit{ApJ}, \textbf{459}, 110.

\bibitem[]{Fukugita,Hogan&Peebles1998} Fukugita, M., Hogan, C. J.,
  and Peebles, P. J. E. (1998). \textit{ApJ}, \textbf{503}, 518.

\bibitem[]{Gonzalez1993} Gonz\'alez, J. J.\ (1993). PhD Thesis,
  University of California, Santa Cruz.

\bibitem[]{Gregg et al.2004} Gregg, M.~D., Ferguson, H.~C., Minniti,
  D., Tanvir, N., and Catchpole, R.\ (2004). \textit{AJ},
  \textbf{127}, 1441.

\bibitem[]{Greggio1997} Greggio, L.\ (1997). \textit{MNRAS},
  \textbf{285}, 151.

\bibitem[]{Grillmair et al.1996} Grillmair, C.~J., Lauer, T.~R.,
  Worthey, G., et al.\ (1996). \textit{AJ}, \textbf{112}, 1975.

\bibitem[]{Harris & Harris2000} Harris, G.~L.~H.~and Harris, W.~E.\ (2000). 
\textit{AJ}, \textbf{120}, 2423.  

\bibitem[]{Harris & Harris2002} Harris, W.~E.~and Harris, G.~L.~H.\ (2002). 
\textit{AJ}, \textbf{123}, 3108.  

\bibitem[]{Humphrey & Buote2006} Humphrey, P.~J.~and Buote, D.~A.\ (2006). 
\textit{ApJ}, \textbf{639}, 136.  

\bibitem[]{Jorgensen et al.2005} J{\o}rgensen, I., Bergmann, M.,
  Davies, R., Barr, J., Takamiya, M., and Crampton,
  D.\ (2005). \textit{AJ}, \textbf{129}, 1249.

\bibitem[]{Kuntschner2000} Kuntschner, H.\ (2000). \textit{MNRAS},
  \textbf{315}, 184.

\bibitem[]{Kuntschner et al.2006} Kuntschner, H., Emsellem, E., Bacon,
  R., et al.\ (2006). \textit{MNRAS}, \textbf{369}, 497.

\bibitem[]{Maraston & Thomas2000} Maraston, C.~and Thomas, D.\ (2000).
  \textit{ApJ}, \textbf{541}, 126.

\bibitem[]{Nelan et al.2005} Nelan, J.~E., Smith, R.~J., Hudson,
  M.~J., Wegner, G.~A., Lucey, J.~R., Moore, S.~A.~W., Quinney, S.~J.,
  and Suntzeff, N.~B.\ (2005). \textit{ApJ}, \textbf{632}, 137.

\bibitem[]{Oconnell1980} O'Connell, R.~W.\ (1980). \textit{ApJ},
  \textbf{236}, 430.

\bibitem[]{Peletier1989} Peletier, R.~F.\ (1989). PhD Thesis,
  Rijksuniversiteit Groningen.

\bibitem[]{Peterson1976} Peterson, R.~C.\ (1976). \textit{ApJ},
  \textbf{210}, L123.

\bibitem[]{Pietrinferni, Cassisi, Salaris, & Castelli2004}
  Pietrinferni, A., Cassisi, S., Salaris, M., and Castelli,
  F.\ (2004). \textit{ApJ}, \textbf{612}, 168.

\bibitem[]{Pietrinferni, Cassisi, Salaris, & Castelli2006}
  Pietrinferni, A., Cassisi, S., Salaris, M., and Castelli,
  F.\ (2006). \textit{ApJ}, \textbf{642}, 797.

\bibitem[]{Rabin1982} Rabin, D.\ (1982). \textit{ApJ}, \textbf{261},
  85.

\bibitem[]{Rejkuba et al.2005} Rejkuba, M., Greggio, L., Harris,
  W.~E., Harris, G.~L.~H., and Peng, E.~W.\ (2005). \textit{ApJ},
  \textbf{631}, 262.

\bibitem[]{Salaris & Weiss1998} Salaris, M.~and Weiss, A.\ (1998). 
\textit{A\&A}, \textbf{335}, 943.  

\bibitem[]{Salaris, Chieffi, & Straniero1993} Salaris, M., Chieffi,
  A., and Straniero, O.\ (1993). \textit{ApJ}, \textbf{414}, 580.

\bibitem[]{Sanchez-Blazquez, Gorgas, Cardiel, & Gonzalez2006}
  S{\'a}nchez-Bl{\'a}zquez, P., Gorgas, J., Cardiel, N., and
  Gonz{\'a}lez, J.~J.\ (2006). \textit{A\&A}, \textbf{457}, 809.

\bibitem[]{Schechter&Dressler1987} Schechter, P. L. and Dressler,
  A. (1987).  \textit{AJ}, \textbf{94}, 563.

\bibitem[]{Schulte-Ladbeck, Drozdovsky, Belfort, & Hopp2003}
  Schulte-Ladbeck, R.~E., Drozdovsky, I.~O., Belfort, M., and Hopp,
  U.\ (2003). \textit{Ap\&SS}, \textbf{284}, 909.

\bibitem[]{Serra&Trager2006} Serra, P. and Trager, S.~C. (2006).
  \textit{MNRAS}, in press.

\bibitem[]{Serven, Worthey, & Briley2005} Serven, J., Worthey, G., and 
Briley, M.~M.\ (2005). \textit{ApJ}, \textbf{627}, 754.  

\bibitem[]{Stasinska, Richer, & McCall1998} Stasi{\'n}ska, G., Richer,
  M.~G., and McCall, M.~L.\ (1998). \textit{A\&A}, \textbf{336}, 667.

\bibitem[]{Stasinska et al.2005} Stasi{\'n}ska, G., V{\'{\i}}lchez,
  J.~M., P{\'e}rez, E., Corradi, R.~L.~M., Mampaso, A., and Magrini,
  L.\ (2005).  \textit{AIPC}, \textbf{804}, 262.

\bibitem[]{Stasinska2006} Stasi{\'n}ska, G., V{\'{\i}}lchez, J.~M.,
  P{\'e}rez, E., Gonzalez Delgado, R.~M., Corradi, R.~L.~M., Mampaso,
  A., and Magrini, L. (2006).  In \textsl{Planetary Nebulae Beyond the
    Milky Way}, ed.\ L. Stanghellini, J. R. Walsh, and N. G. Douglas
  (Springer Verlag, Berlin), 234.

\bibitem[]{Thomas, Maraston, & Bender2003} Thomas, D., Maraston, C.,
  and Bender, R.\ (2003). \textit{MNRAS}, \textbf{339}, 897.

\bibitem[]{Thomas, Maraston, Bender, & Mendes de Oliveira2005} Thomas,
  D., Maraston, C., Bender, R., and Mendes de Oliveira,
  C.\ (2005). \textit{ApJ}, \textbf{621}, 673.

\bibitem[]{Tinsley1980} Tinsley, B.~M.\ (1980). \textit{FCPh},
  \textbf{5}, 287.

\bibitem[]{Trager2004} Trager, S.~C.\ (2004). In \textit{Origin and
  Evolution of the Elements}, ed.\ A. McWilliam and M. Rauch
  (Cambridge University Press, Cambridge), 388.

\bibitem[]{TFD06} Trager, S.~C., Faber, S.~M., and Dressler, A.
  (2006), in preparation

\bibitem[]{Trager, Faber, Worthey, & Gonzalez2000a} Trager,
  S. C., Faber, S.  M., Worthey, G., and Gonz\'alez,
  J. J.\ (2000a). \textit{AJ}, \textbf{119}, 1645.

\bibitem[]{Trager, Faber, Worthey, & Gonzalez2000b} Trager,
  S. C., Faber, S.  M., Worthey, G., and Gonz\'alez,
  J. J.\ (2000b). \textit{AJ}, \textbf{120}, 165.

\bibitem[]{Trager, Worthey, Faber, & Dressler2005} Trager, S.~C.,
  Worthey, G., Faber, S.~M., and Dressler, A.\ (2005). \textit{MNRAS},
  \textbf{362}, 2.

\bibitem[]{VandenBerg et al.2000} VandenBerg, D.~A., Swenson, F.~J., 
Rogers, F.~J., Iglesias, C.~A., and Alexander, D.~R.\ (2000). \textit{ApJ}, 
\textbf{532}, 430.  

\bibitem[]{Walsh, Walton, Jacoby, & Peletier1999} Walsh, J.~R.,
  Walton, N.~A., Jacoby, G.~H., and Peletier,
  R.~F.\ (1999). \textit{A\&A}, \textbf{346}, 753.

\bibitem[]{Walsh, Jacoby, Peletier, & Walton2006} Walsh, J.~R.,
  Jacoby, G., Peletier, R., and Walton, N.~A.\ (2006). In
  \textsl{Planetary Nebulae Beyond the Milky Way},
  ed.\ L. Stanghellini, J. R. Walsh, and N. G. Douglas (Springer
  Verlag, Berlin), 262.

\bibitem[]{Worthey1994} Worthey, G.\ (1994). \textit{ApJS},
  \textbf{95}, 107.

\bibitem[]{Worthey2004} Worthey, G.\ (2004). \textit{AJ},
  \textbf{128}, 2826.

\bibitem[]{Worthey, Dorman, & Jones1996} Worthey, G., Dorman, B., and
  Jones, L.~A.\ (1996). \textit{AJ}, \textbf{112}, 948.

\bibitem[]{Worthey, Faber, & Gonzalez1992} Worthey, G., Faber, S.~M.,
  and Gonz\'alez, J.~J.  (1992). \textit{ApJ}, \textbf{398}, 69.

\end{thebibliography}
\end{document}